\documentclass[aps,prl,twocolumn,showpacs,superscriptaddress]{revtex4-2}
\usepackage{graphicx}  
\usepackage{dcolumn}   
\usepackage{bm}        
\usepackage{amssymb}   
\usepackage{float}
\usepackage{amsmath} 
\usepackage{xcolor}
\usepackage{ulem} 
\usepackage{xr}
\usepackage{comment}

\usepackage{soul}

\newcommand{\reponse}[1]{\textcolor{black}{#1}}

\begin{document}

\title{Tunable Thin Elasto-Drops}
\author{A.~Eddi}
\author{S.~Perrard}
\author{J.~Zhang}
\email{jishen.zhang@espci.fr}
\affiliation{PMMH, CNRS, ESPCI Paris, Université PSL, Sorbonne Université, Université de Paris, F-75005, Paris, France}
\date{\today}

\begin{abstract}
We present an experimental method to fabricate centimetric thin elastic capsules with highly uniform thickness and negligible bending stiffness using silicone elastomers. In our experiments, the capsules thickness is tunable at fabrication, while internal pressure and hoop (circumferential) stress are adjustable via hydrostatic inflation once the capsules are filled and immersed in water. Capsules mechanics are probed through hydro-elastic waves generated by weak mechanical perturbations at the capsule interface. By analyzing the surface wave dynamics in the Fourier domain, we extract the in-plane stress and demonstrate that the hydro-elastic waves are exclusively governed by hoop stress. This \reponse{provides a controllable macroscopic analogue of liquid drops} characterised by an effective surface tension, allowing the capsules to be modeled as large-scale ``elasto-drops'' with an inflation and thickness tunable effective surface tension. \reponse{In this limit, bending stiffness is negligible over the experimentally relevant wavelengths, so that the shell dynamics are governed primarily by in-plane tension.} Our work demonstrates that elasto-drops serve as a robust model system for parametric studies of large-scale \reponse{analogues of} liquid drops with experimentally adjustable surface tension.
\end{abstract}

\maketitle

\section{Introduction}

Soft particles such as droplets and bubbles deform when subjected to an external flow or during impacts. The deformation in turn governs their dynamics, including impact spreading~\cite{clanet2004maximal}, vortex shedding~\cite{mougin2001path}, as well as oscillatory rise dynamics under double confinement~\cite{pavlov2022oscillating}. These behaviours are typically quantified using dimensionless numbers such as the Weber number for droplet impacts or drops in flows, which compares inertia and surface tension, the Bond number for buoyancy-driven bubble deformation, which compares gravitational forces and surface tension, or the Cauchy number for soft particles, which compares inertia and elastic modulus. For droplets and bubbles, the accessible ranges of Weber or Bond number are intrinsically limited by surface tension $\gamma$, whose values are set by intermolecular forces. While surface tension can span more than an order of magnitude across liquid families (e.g. 10–487 mN/m for common liquids at 20°C)~\cite{adamson1967physical}, altering $\gamma$ inevitably modifies other fluid properties such as density and viscosity. As a result, the capillary length remains millimetric for nearly all standard liquids. In addition, the presence of surfactants either naturally or introduced voluntarily can significantly alter the surface tension of fluid pairs~\cite{eastoe2000dynamic}, which further reduces the effective controllability of the interfacial properties.

\reponse{An alternative strategy consists in replacing the fluid interface by a thin elastic shell whose mechanical response can be tuned through its geometry and internal pressure. The mechanics of thin elastic shells and capsules has been extensively investigated over the past decades in contexts ranging from biological membranes to synthetic capsules and pressurized shells. When the shell thickness is sufficiently small, bending contributions become negligible and the response is governed predominantly by in-plane stretching and membrane tension. In this limit, the global mechanical behaviour of a capsule is largely controlled by the membrane tension, which can be adjusted via internal pressure or imposed deformation.}

\reponse{Compression experiments on liquid-filled elastic capsules have shown that, in the thin-shell limit where bending rigidity is negligible, the restoring force is governed predominantly by in-plane membrane tension~\cite{risso2004compression}. For shells subjected to an imposed internal overpressure, this tension is set directly by the pressure, which generates an approximately uniform in-plane stress and thereby controls the effective stiffness of the shell~\cite{vella2012indentation}.}

\reponse{Together, these results show that sufficiently thin elastic capsules can be driven into a regime where their mechanical response is dominated by membrane tension. This offers a convenient route to explore soft-particle dynamics in regimes that are difficult to access with conventional fluid interfaces, for which surface tension is fixed by molecular interactions.}

\reponse{Recent experimental advances in the fabrication of hollow elastic shells using elastomers such as VPS (vinyl polysiloxane) or PDMS (polydimethylsiloxane) have enabled controlled production of elastic spheres with prescribed geometry and mechanical properties~\cite{lee2016fabrication, nasto2013localization}. Such elastic capsules have been employed as model systems in various contexts, including droplet impact dynamics on solid walls~\cite{jambon2020deformation}, shear-induced deformation and breakup of microcapsules~\cite{haner2020deformation,joung2020synthetic}, and flow–structure interactions of soft particles. In particular, Jambon-Puillet \textit{et al.}~\cite{jambon2020deformation} showed that elastic water capsules can extend the regime of droplet spreading at impact over Weber numbers an order of magnitude larger than those accessible with weakly viscous droplets~\cite{liu2025viscous}, opening a promising route to probe new dynamical regimes for large-scale soft particle deformations. However, previously studied macroscopic elastic capsules, although highly deformable and successfully described within a membrane framework~\cite{jambon2020deformation}, generally retain a finite bending stiffness that may influence their response depending on the deformation mode and the excited length scales. Accessing the asymptotic thin-shell limit in which bending rigidity becomes negligible over the full range of dynamically relevant wavelengths remains experimentally challenging. Such an ultra-thin regime would enable a closer analogy with capillary interfaces governed purely by an effective surface tension set by in-plane stress.}

\reponse{In this tension-dominated limit, thin elastic capsules provide macroscopic systems whose dynamics are governed primarily by in-plane membrane stress, in close analogy with capillary interfaces. In the following, we refer to these thin elastic shells as ``elasto-drops'' to emphasize this analogy: although they are hollow elastic capsules rather than fluid droplets, their effective surface tension can be tuned through inflation and thickness, enabling controlled studies of drop-like dynamics with independently adjustable mechanical properties.}

In this work, we introduce a new method to fabricate centimetric thin elastic capsules with controlled thickness and tunable effective tension. Our approach builds on previous studies showing that elastic membranes strongly modify surface-wave dynamics, as demonstrated in investigations of hydro-elastic waves on floating elastic sheets, ranging from ice-sheet analogues~\cite{auvity2025wave,schulkes1987waves} and elastic wave turbulence~\cite{vernet2025thermodynamics,deike2013nonlinear}, to optical and solid-state physics analogues~\cite{domino2018dispersion,domino2020artificial,doudic2024measuring}. For an elastic sheet of thickness $h$, Young’s modulus $E$, and Poisson ratio $\nu$, hydro-elastic wave dynamics in the small-amplitude limit are obtained by coupling the linearized Föppl–von K\'arm\'an equations describing the in-plane tension and bending response of the sheet with the Bernoulli equation for an inviscid, irrotational fluid. In the deep-water limit, this coupling yields the standard hydro-elastic dispersion relation $\omega(k)$~\cite{schulkes1987waves}

\begin{equation}
    \omega^2=gk + \frac{T}{\rho} k^3 + \frac{D}{\rho} k^5, 
    \quad D=\frac{Eh^3}{12(1-\nu^2)},
\end{equation}

where $g$ is gravity, $T$ the in-plane tension and $D$ the bending modulus. For a closed elastic capsule filled with water and fully immersed in water, the gravitational contribution vanishes and the relevant in-plane tension reduces to the hoop stress of the spherical membrane. By measuring the dispersion relation of waves propagating along the capsule surface through controlled mechanical excitation, we directly extract this hoop stress and thus obtain the capsule’s effective tension in a simple, non-intrusive manner.

\reponse{This dynamic measurement provides a complementary approach to the static mechanical tests commonly used to characterize elastic shells, such as compression or indentation experiments~\cite{risso2004compression,vella2012indentation}. Whereas static tests probe the local force–displacement response of the shell, the present wave-based method directly measures the global membrane tension governing the capsule dynamics. This enables a direct characterization of tension-dominated shells in a fully submerged configuration and provides a convenient route to establish a quantitative analogy with capillary interfaces possessing a tunable effective surface tension.}

\section{Materials \& methods}

\subsection{Elasto-drop fabrication}

\begin{figure}[h]
  \centering
  \includegraphics[width=0.9\linewidth]{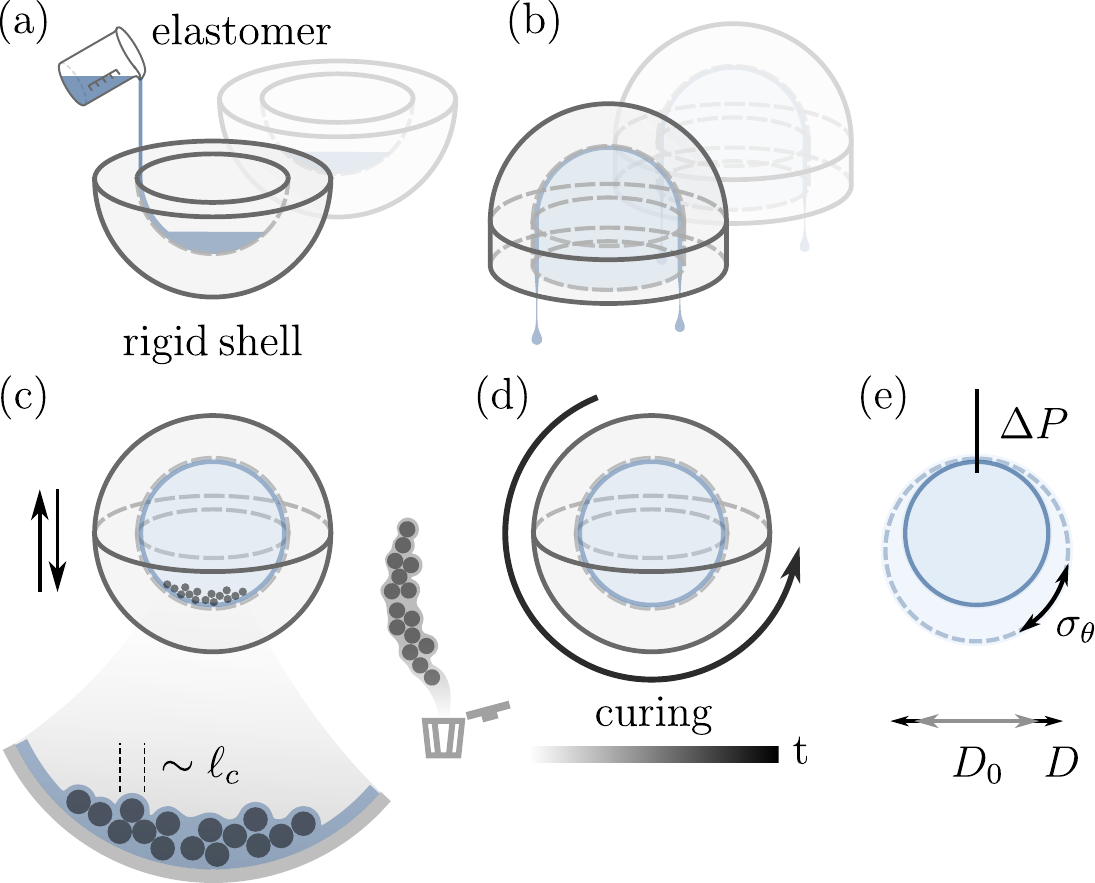}
  \caption{\label{fig:fabrication} Fabrication process of the elasto-drop. a–e: Sequence of coating, draining, ball-assisted thinning, curing, demolding and liquid filling.}
\end{figure}

The elastic shell of uniform thickness was fabricated using a viscous coating technique adapted from established methods~\cite{jambon2020deformation, lee2016fabrication}. The fabrication protocol is illustrated in Fig.~\ref{fig:fabrication}. \reponse{A two-part PDMS-based silicone elastomer (Ecoflex 00-30) is mixed at equal weight ratio of base and catalyst. The resulting polymer solution has a pot life of approximately 45~min and cures within about 1~h at room temperature, allowing sufficient time for thickness homogenization prior to cross-linking.}

\reponse{The liquid polymer is first deposited into each hemisphere of a spherical plastic mold and spread over the inner surface. The hemispheres are placed bottom-down on supporting rings to allow excess liquid to drain without accumulating at the rim. To further reduce the film thickness, millimetric plastic balls are introduced into the hemispheres, which are then closed and vigorously shaken for several seconds in all directions. This rapid multidirectional shaking ensures that the balls roll over the entire inner surface and become fully coated with liquid elastomer, thereby removing excess material. The operation must be performed as quickly as possible so that the liquid has time to redistribute and smooth out before significant cross-linking occurs. The achievable thickness can be tuned by adjusting the initial deposited volume and the intensity of the ball-assisted thinning step, with a larger number of beads leading to thinner shells. After removal of the balls, the mold is sealed and mounted on a planetary rotating platform during curing; capillary forces progressively smooth out residual surface imprints.}

\reponse{Once fully cured, the elastic shell is demolded and filled with water through a needle puncture. The puncture site is sealed with a small drop of fast-curing elastomer. The resulting elasto-drops are highly deformable owing to the low Young's modulus of Ecoflex 00-30 ($E=62.5\pm37.5$~kPa), independently measured by tensile tests on strip samples.}

\reponse{We fabricate capsules of initial external diameter $D_0=42$~mm with three representative thicknesses. The mean shell thickness is first determined from the total capsule mass and elastomer density, yielding an weight-average thickness $\langle h_0\rangle$. Independent measurements are obtained by white-light interferometry on excised membrane strips of width $\approx 2$~mm, providing local thickness profiles along a meridian. For the three capsules studied here, the mass-based thicknesses are 59.2, 125.5 and 185.5~$\mu$m, while the corresponding arithmetic averages from interferometry are 60.2$\pm9.0$, 124.3$\pm25.7$ and 164.7$\pm16.4$~$\mu$m. The two estimates are in close agreement for the thinner shells but differ more significantly for the thickest capsule, indicating large-scale thickness heterogeneities.}

\begin{figure}[h!]
  \centering
  \includegraphics[width=0.92\linewidth]{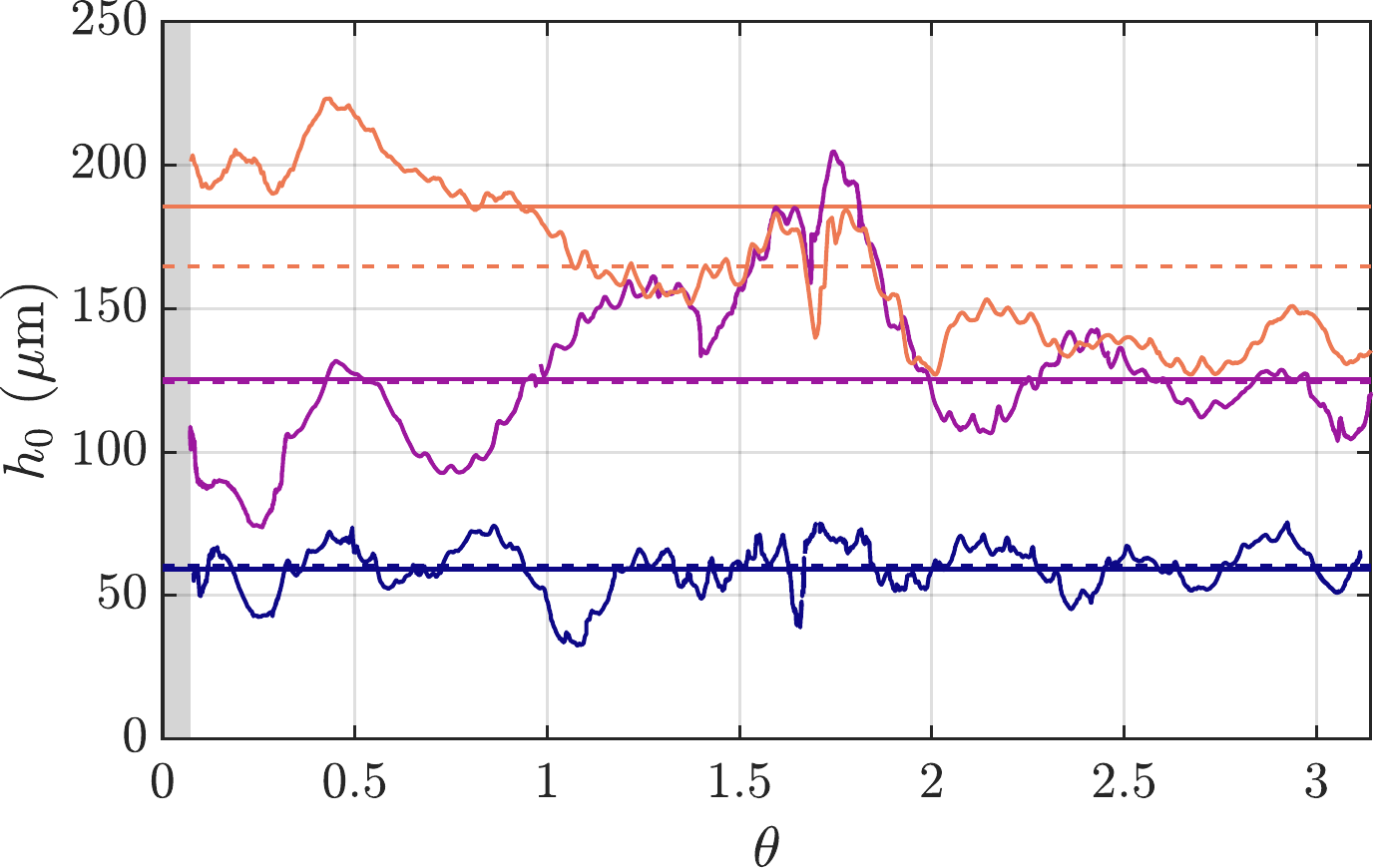}
  \caption{\label{fig:thickness} \reponse{Meridional thickness profile $h_0(\theta)$ measured by white-light interferometry for three elasto-drops. Colors correspond to the averaged thickness determined from capsule mass: blue: $\langle h_0\rangle=59.2~\mathrm{\mu m}$, violet: $125.5~\mathrm{\mu m}$, and orange: $185.5~\mathrm{\mu m}$. Solid lines indicate the mass-based average thickness, while dashed lines indicate the arithmetic mean thickness obtained from the interferometry measurements along the meridian. Shaded bar: mount joint (see Fig.~\ref{fig:Sketch}(b)).}}
\end{figure}

\reponse{Figure~\ref{fig:thickness} shows the meridional thickness profile $h_0(\theta)$ for the three capsules, where $\theta$ refers to the zenith angle. Thickness uniformity is quantified by the coefficient of variation $\mathrm{CV}=\sigma_{h_0}/\langle h_0\rangle$, where $\sigma_{h_0}$ is the spatial standard deviation along the meridian. 
The measured values are $\mathrm{CV}=15\%$, $21\%$ and $12\%$ for increasing mean thickness. 
Although spatial variations are not negligible, the thickness remains sufficiently homogeneous for the membrane to sustain an approximately uniform hoop stress under inflation.}

\reponse{In the following, we use exclusively the thickness obtained from capsule mass. This definition provides the true overall thickness of the membrane and is therefore the appropriate quantity for converting the measured in-plane tension into a representative hoop stress. Because the hydro-elastic waves probe the global membrane tension, the resulting stress--strain relation must be based on the total elastic area of the shell rather than on a local or meridional average thickness, which can vary depending on where the membrane is sampled on the capsule.}

\reponse{\subsection{Fabrication limits}}

\reponse{In practice, both the minimum achievable thickness and its spatial uniformity are constrained by viscous--capillary relaxation during reticulation and by geometric defaults of the mold.}

\reponse{During the ball-assisted thinning stage, the rolling beads create transient grooves and local thickness variations in the liquid elastomer layer. After shaking, these perturbations progressively relax under the action of surface tension. The associated capillary healing time depends on the viscosity of the uncured elastomer and on the film thickness. For a groove of lateral size $L_b\sim\sqrt{d_bh}$, where $d_b$ is the ball diameter and $h$ the local film thickness, the characteristic relaxation time derived from the thin film equation~\cite{stillwagon1988fundamentals, mcgraw2012self} can be estimated as}
\begin{equation}
    t_h \sim \frac{3\mu}{\gamma}\frac{d_b^2}{h},
    \label{eq:t_h}
\end{equation}
\reponse{where $\mu$ and $\gamma$ denote the dynamic viscosity and surface tension of the liquid elastomer. For freshly mixed Ecoflex 00-30 ($\mu\approx3~\mathrm{Pa\cdot s}$, $\gamma\approx20$~mN/m) and millimetric balls ($d\approx3$~mm), this estimate yields healing times ranging from a few tens of seconds for $h\sim150~\mu$m to several minutes when the thickness approaches $h\sim50~\mu$m.}

\reponse{As the elastomer progressively cross-links, its viscosity increases and eventually stops the capillary smoothing process. When the healing time becomes comparable to the curing time, residual thickness heterogeneities may persist in the final shell. This competition between capillary smoothing and cross-linking therefore sets a first practical constraint on the minimum uniform thickness that can be achieved.}

\reponse{For thicker coatings, an additional source of inhomogeneity arises from gravity-driven drainage prior to curing. When the liquid elastomer layer becomes sufficiently thick, its own weight can induce slow downward flow along the inner surface of the mold, similar to a falling-film regime. This redistribution occurs over timescales comparable to the curing time and can generate large-scale thickness gradients before cross-linking arrests the flow, contributing to the increased dispersion observed for thicker shells. During curing, the mold is mounted on a planetary rotating platform to promote thickness homogenization. The resulting thickness distribution therefore reflects a competition between gravity-driven drainage along the mold and rotational redistribution induced by the platform motion. If the rotation is too slow, gravitational flow dominates and leads to accumulation in the lower regions of the mold, an optimal rotation rate must be identified to minimize large-scale thickness gradients and ensure the most uniform shell possible.}

\reponse{A second limitation arises from the finite gap at the junction between the two hemispherical molds. Excess liquid tends to accumulate along the rim of each hemisphere during draining and shaking, and when the mold is closed this liquid cannot fully redistribute across the small interfacial gap. As a result, for very thin coatings the remaining fluid preferentially collects near the edges rather than forming a continuous uniform film across the junction. The combined effects of incomplete capillary smoothing and edge accumulation therefore set a practical lower bound of order $50\,\mu$m for obtaining homogeneous shells in our fabrication protocol.}

\subsection{Experimental setup}\label{sec:exp}

\begin{figure}[h!]
  \centering
  \includegraphics[width=0.95\linewidth]{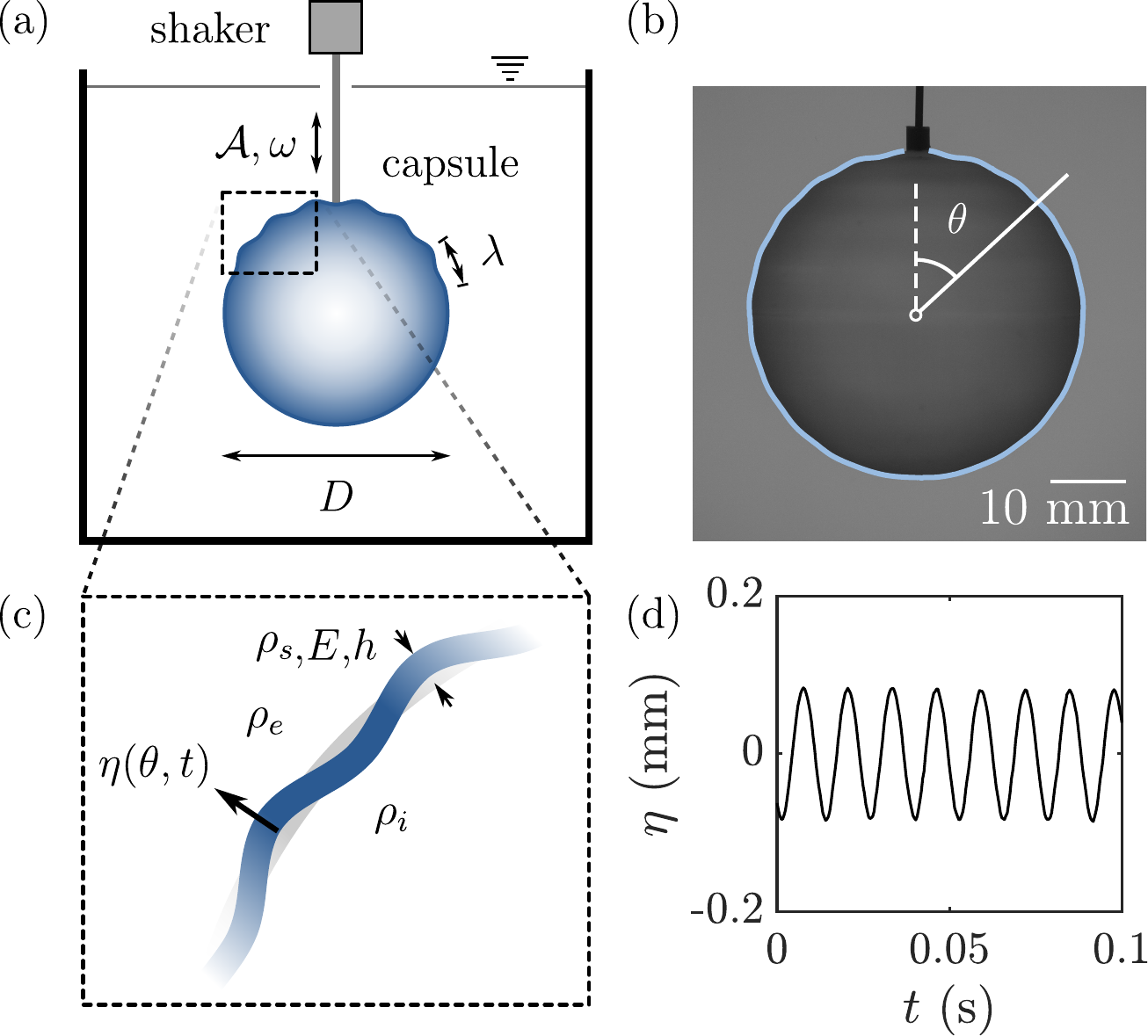}
  \caption{\label{fig:Sketch} a: Schematics of the hydro-elastic wave propagation on the surface of a pre-stretched elasto-drop. b: Example of the elasto-drop's surface edge detection, for an elasto-drop of initial thickness $\reponse{\langle h_0\rangle}=125.5~\mathrm{\mu m}$, inflated at a strain $(R-R_0)/R=4~\%$, under a periodic forcing frequency $f_s=77.7$ Hz. The blue solid line represents the elasto-drop's detected edge. The zenith angle $\theta$ varying from 0 to $2\pi$ in the clockwise direction is defined with the vertical dashed line passing through the center and a solid line. c: Physical properties of the elastic shell and the surrounding fluids. The surface elevation is in the radial direction of the initially unperturbed sphere. d: Time variation of the surface elevation $\eta(\theta=0.82, t)$, obtained based on a classical edge detection by image intensity gradient.}
\end{figure}

The elasto-drop, filled up with water, is placed at the center of a water tank of dimension $250\times 250 \times 200~\mathrm{mm^3}$, as is shown in Figure~\ref{fig:Sketch}(a). A needle of 1 mm in diameter and 117 mm in length is used to hold the elasto-drop in place. The top of the shell is pierced by the lower tip of a needle and sealed with elastomer, allowing for rapid variation of the liquid volume inside the elasto-drop by injecting or extracting water with a syringe. We are thus able to control the hydrostatic pressure $\Delta P$ inside the shell. This pressure due to the inflation can be further associated to the hoop stress $\sigma_\theta$ by $\Delta P= 2h/R \sigma_\theta$~\cite{ross2021mechanics}~\reponse{(Fig.~\ref{fig:Sketch}(b))}. 

For spherical elasto-drop inflation, the shell undergoes an a uniform and isotropic deformation (equibiaxial), the tangential strain in the circumferential direction can be simply associated by the radius variation with respect to the initial radius $\varepsilon = (R-R_0)/R$. In our experimental investigations, the elasto-drops were inflated at ten discrete strain increments, spanning a tangential strain from 0 to approximately $\varepsilon\approx 30\%$. Under this weak deformation regime, we expect that the stress remains linearly proportional to the strain $\sigma_\theta=E/(1-\nu)\varepsilon$~\cite{landau2012theory}.

The needle is firmly mounted to an electromagnetic shaker (LDS V200) that allows to generate mechanical perturbations with controlled frequencies and amplitudes (Fig.~\ref{fig:Sketch}a). Via the needle, the perturbations are directly transmitted to the top of the elasto-drop's surface and elastic waves propagate along the elasto-drop's surface. The wave propagation being axisymmetric about the vertical axis, the elastic wave can be directly associated to the surface elevation $\eta (\theta, t)$, where $\theta$ is the zenith angle~(Fig.~\ref{fig:Sketch}b). To capture the elasto-drop's edge displacement, LED-backlighted images of the elasto-drop contour have been recorded using a high-speed camera (Phantom v1840) equipped with a Samyang f2.8/100 optical lens leading to a pixel size of $59~\mathrm{\mu m}$~(Fig.~\ref{fig:Sketch}b). 

To cross validate our elasto-drop's mechanical property extraction from wave measurements, two types of stroking signals were used and are subsequently referred as the \textit{harmonic} and \textit{impulse} forcing. The \textit{harmonic} forcing mode has an analytical expression $A(t)=A_0 \cos (2 \pi f_s t)$. For each given strain of elasto-drop, 20 points were recorded with the forcing frequency $f_s$ logarithmically spaced between 10 Hz and 345 Hz. For the \textit{impulse} mode, the capillary tip moves downward from the initial position at rest before reaching a maximum amplitude, then lifts back upward to the starting position. The total time duration of the stroke is around 10 ms.


The recording sampling rate for the \textit{harmonic} mode was varied for each wave stroke frequency to ensure a minimum of 16 images per wave time period $1/f_s$, for a recording duration of at least 100 wave periods. For the \textit{impulse} mode, each recording lasts 2 seconds with a sampling frequency $f_s=3000$ Hz.

We extract and analyse the elasto-drop's spatiotemporal edge elevation in polar coordinates $\eta(\theta,t)$, consistent with wave propagation on a spherical surface. To do so, we used a matlab built-in function \textit{improfile} to sample the image intensity values along a line segment passing through the center of the shell, starting from the vertical top origin where $\theta=0$ and sweeping clockwise to $\theta=2\pi$, with a step of $\delta \theta=0.5^\circ$. At a given angle, the associated elevation is obtained with a classical 1D edge detection method, based on spatial gradient of the intensity variation with a subpixel resolution. A snapshot image of the elasto-drop under \textit{harmonic} forcing at $f_s=77.7$ Hz is shown in Figure~\ref{fig:Sketch}(b), the elasto-drop has a strain $(R-R_0)/R_0=4\%$. The blue curve corresponds to the extracted edge deformation $\eta(\theta,t)$. The wavy deformation can be seen to be symmetrical about the vertical axis. Due to the attachment mount of the elasto-drop to the capillary tip, the total angular range of the elastic shell is slightly restricted by $4.1^\circ$ from each end. Figure~\ref{fig:Sketch}(d) illustrates a time evolution of the local edge displacement in periodic motion at the zenith angle $\theta=47^\circ$.


\section{Results \& discussion}

\begin{figure}[ht]
  \centering
  \includegraphics[width=0.9\linewidth]{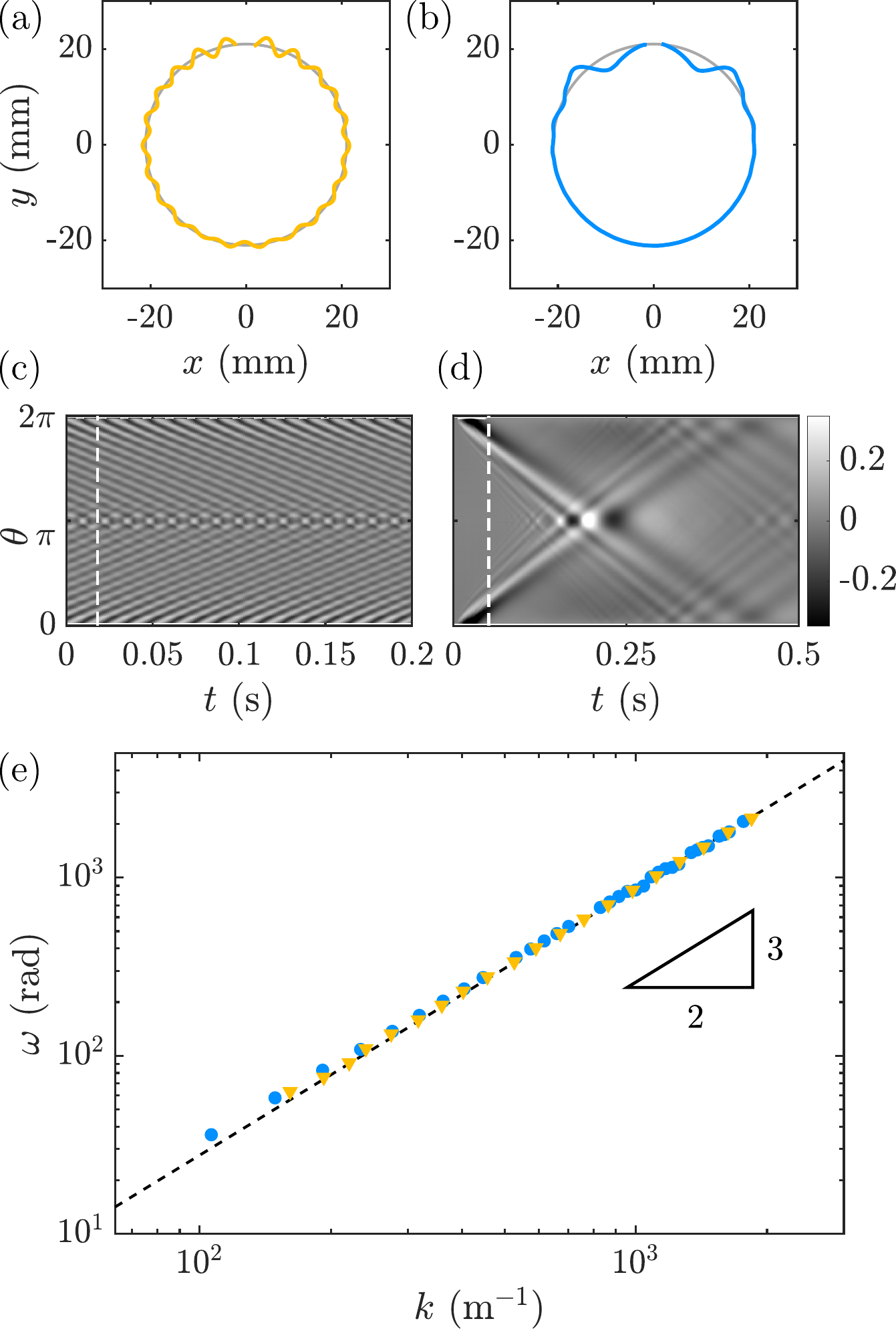}
  \caption{\label{fig:instantaneous_polaire} a,b: Instantaneous shapes of the elasto-drop surface under a-\textit{harmonic} and b-\textit{impulse} forcing, radially amplified by a factor of ten for better visibility. c,d: Spatiotemporal diagram of the surface amplitude $\eta$ as a function of the zenith angle $\theta$. The \textit{harmonic} forcing frequency reads $f_s=77.7$ Hz. e: Dispersion relation $\omega(k)$ extracted from the frequency-wavenumber domain of the surface amplitude $\hat{\theta}(\omega, k)$ for \textit{harmonic} (yellow) and \textit{impulse} (blue). The dashed lines indicate the power-law fit $\omega=[T k^3/(2\rho)]^{1/2}$.}
\end{figure}

Fig.~\ref{fig:instantaneous_polaire}(a,b) shows snapshots of the typical elasto-drop's surface elevation under two forcing modes, measured at an inflation strain $\varepsilon=4\%$. The corresponding spatio-temporal amplitudes $\eta(\theta, t)$ are illustrated in Fig.~\ref{fig:instantaneous_polaire}(c,d), where the dashed lines indicate the time position of the above shown instantaneous forms. We see on the diagrams that under both forcing modes, waves are emitted from the top ($\theta=0$, $2\pi$), propagate from both ends towards the bottom of the shell, forming a symmetry about the vertical axis ($\theta=\pi$). It is noticeable that for waves under \textit{impulse} forcing (b,d), a single impulsive impact on the elasto-drop generates a large-band of dispersive waves in a broad band of wavelengths.

We now focus on the dispersion relation analyses of the wave dynamics. To this end, we first perform a coordinate transformation from the zenith angle $\theta$ to the arc length $x = R\theta$, which describes wave propagation in the in-plane (tangential) direction, where $R$ is the unperturbed radius of the elasto-drop. From the spatiotemporal signal $\eta(x, t)$, we compute its Fourier transform $\hat{\eta}(\omega, k)$ in frequency and wavenumber space. For \textit{harmonic} forcing, we identify the dominant wavenumber $k_m$ corresponding to the maximum spectral amplitude at each stroke frequency $f_s$. The extracted pairs $(f_s, k_m)$ are then displayed in Fig.~\ref{fig:instantaneous_polaire}(e) as yellow triangles in log-log scale. The same procedure can be applied to waves under \textit{impulse} forcing; however, in this case multiple frequencies and wavenumbers can be simultaneously excited, since the wave field is inherently multi-spectral. The extracted pairs are illustrated with blue dots. When comparing the extracted dispersion relations from the two forcing modes, we find an excellent agreement. In addition, we plot the theoretical power law $\omega \propto k^{3/2}$, which spans more than a decade in both frequency and wavenumber, consistent with the expected behaviour of tension-dominated waves. We then fit the data with a power law (dashed line) to extract the corresponding tension $T$.

In our case, in absence of the free surface, the elasto-drop encapsulating water is fully submerged in a water tank, the gravity term $\omega^2\sim gk$ vanishes. Moreover, the consistency of our data with the power law $k^{3/2}$ suggests also that the flexural term vanishes. The dispersion relation of the waves then becomes $\omega^2=T/(\rho_i+\rho_e) k^3$, where both external and internal fluids are water in this case, $\rho=\rho_e=\rho_i$. 

\begin{figure}[h]
  \centering
  \includegraphics[width=0.95\linewidth]{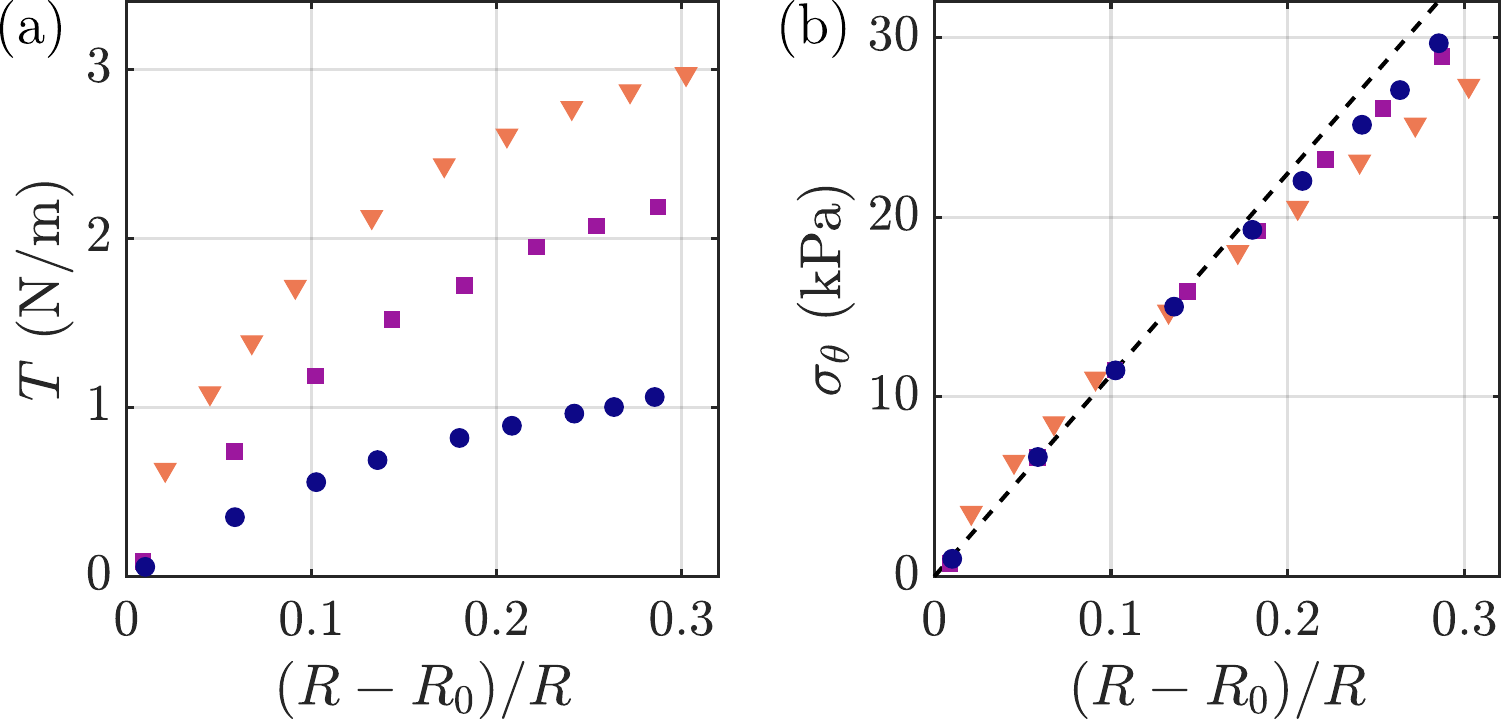}
  \caption{\label{fig:tension_strain} a: Extracted tension of the elasto-drops, for increasing inflation, as a function of the in-plane strain. \reponse{B}lue: $\reponse{\langle h_0\rangle}=59.2~\mathrm{\mu m}$, violet: $125.5~\mathrm{\mu m}$ and orange: $185.5~\mathrm{\mu m}$. b: Tangential stress $\sigma_\theta$ as a function of the strain.}
\end{figure}

Fig.~\ref{fig:tension_strain}(a) shows  the extracted tension $T$ as a function of the principle in-plane strain of the inflated elasto-drop, for three different initial shell thicknesses (blue: $\reponse{\langle h_0\rangle}=59.2~\mathrm{\mu m}$, violet: $125.5~\mathrm{\mu m}$ and orange: $185.5~\mathrm{\mu m}$). We see that increasing either the inflation or the initial thickness of the elasto-drop lead to an increase \reponse{in the membrane} tension.

We now divide the extracted tension by the corresponding shell thickness $h$ at each strain to obtain the in-plane stress $\sigma_\theta$, as shown in Fig.~\ref{fig:tension_strain}(b). We find that all data collapse on a master line following $\sigma_\theta = E_T/(1-\nu) \varepsilon$, where $E_T=56$ kPa is the extracted Young's modulus. This value is in good agreement with different works using the same commercial polymer \cite{lanoy2020dirac,vaicekauskaite2020mapping,
delory2022soft, delory2023guided, delory2024viscoelastic, janardhana2025comprehensive}. \reponse{The agreement confirms that the dispersion-based approach provides a direct measurement of the global membrane tension without mechanical contact.}

As for the flexural wave, a typical transition wavenumber $k_{TD}=\sqrt{T/D}$ between the flexion and tension waves can be obtained by balancing both terms $(T/2\rho) k^3=(D/2\rho) k^5$. With a typical extracted values \reponse{of the} tension $T=1$ N/m, we find a cutoff wavenumber and frequency $k_{TD}\approx 4.0\times 10^{3}~\mathrm{m^{-1}}$, $f_{TD} \approx 1.3\times 10^{3}$ Hz for the thickest tested elasto-drop ($\reponse{\langle h_0\rangle}=185.5~\mathrm{\mu m}$) and $k_{TD}\approx 2.2\times 10^{4}~\mathrm{m^{-1}}$, $f_{TD} \approx 1.7\times 10^{4}$ Hz for the thinnest one ($\reponse{\langle h_0\rangle}=59.2~\mathrm{\mu m}$). \reponse{The resulting cutoff wavenumbers lie well above the experimentally accessible range ($k<10^3$ m$^{-1}$), confirming that all measured waves occur in the pure tension regime. Consequently, the elasto-drop dynamics are governed entirely by membrane tension over the full frequency range explored here. This strong separation of scales ensures that the elasto-drop dynamics remain} dominated by tension waves in the typical frequency range of interest of most potential applications such as interaction with water waves or turbulence. For homogeneous isotropic turbulent (HIT) water flows, a recent review comparing the flow characteristics of major HIT facilities reports that the Kolmogorov length scale $\eta$ typically falls within 18-620 $\mathrm{\mu m}$~\cite{beckedorff2025jet}. This yields a corresponding eddi turnover frequency $f_\eta \equiv 1/t_c \sim \varepsilon^{1/3}\eta^{-2/3}$ that lies between 4-446 Hz, where $\varepsilon$ refers to the turbulent kinetic energy dissipation rate. Over this frequency range, capsule deformations in our experiments \reponse{are} entirely governed by surface traction, with other mechanical modes being negligible. 

The analogy with liquid drops meet several limitations due to the fundamental substantial difference between a liquid-liquid interface and a solid membrane. For an elasto-drop, the effective surface tension depends on the strain itself, and is therefore not conserved during deformation. This can lead to significant difference between a drop and an elasto-drop dynamics as soon as the deformation exceeds the linear order. Moreover, at large deformations, shape instability may occur, such as the emergence of wrinkles in the presence of local membrane compression~\cite{yariv2025large, barthes2009capsule, kantsler2007vesicle}.

\section*{Conclusions}

\reponse{We have presented a method for fabricating centimetric ultra-thin elastic capsules with controlled thickness and tunable in-plane tension. By combining controlled inflation with hydro-elastic wave measurements, we obtain a direct and non-invasive characterization of the global membrane tension governing their dynamics.}

\reponse{The observed behaviour is consistent with classical shell mechanics, in which sufficiently thin pressurized shells respond primarily through in-plane membrane tension. In the regime of small, linear deformations explored here, the wave dynamics are governed by this tension, while bending contributions remain negligible.}

\reponse{In this limit, these ultra-thin capsules provide a controlled macroscopic system for investigating tension-dominated interface dynamics with independently tunable geometry and mechanical properties. These highly deformable objects offer a controlled way to decouple size, deformability, fluid density and viscosity, and open the door to the future design of fully tunable viscoelastic particles.}

\section*{Acknowledgements}
\reponse{We thank Pierre Chantelot, Fabrice Lemoult and Gauthier Verhille for fruitful discussions and Virgile Thiévenaz, Michael Berhanu, Antoine Chateauminois for their help in shell thickness measurements.} This work was supported by the Agence Nationale de la Recherche with grants ANR Lascaturb (Grant No. ANR-23-CE30-0043) and ANR TransWaves (Grant No. ANR-24-CE51-3840).




\bibliography{capsules.bib} 
\bibliographystyle{rsc} 

\end{document}